\documentclass[aps,prd,preprint,superscriptaddress]{revtex4}

\usepackage{latexsym}
\usepackage{amsmath,amssymb}
\usepackage{graphicx}
\usepackage{subfigure}

\newcommand{\re}{\text{Re}}
\begin{document}


\title{Effective potentials in the Lifshitz scalar field theory}

\author{Myungseok Eune}
\email[]{younms@sogang.ac.kr}
\affiliation{Research Institute for Basic Science, Sogang University, Seoul, 121-742, Korea}

\author{Wontae Kim}
\email[]{wtkim@sogang.ac.kr}
\affiliation{Research Institute for Basic Science, Sogang University, Seoul, 121-742, Korea}
\affiliation{Center for Quantum Spacetime, Sogang University, Seoul 121-742, Korea}
\affiliation{Department of Physics, Sogang University, Seoul 121-742, Korea}

\author{Edwin J. Son}
\email[]{eddy@sogang.ac.kr}
\affiliation{Center for Quantum Spacetime, Sogang University, Seoul 121-742, Korea}

\date{\today}

\begin{abstract}
  We study the one-loop effective potentials of the four-dimensional
  Lifshitz scalar field theory with the particular anisotropic scaling
  $z=2$, and the mass and the coupling constants renormalization are performed whereas the finite counterterm is just needed for
the highest order of the coupling because of the mild UV divergence. Finally, we investigate 
  whether the critical temperature for the symmetry breaking can exist or not in this
  approximation. 
\end{abstract}


\maketitle

\section{Introduction}
\label{sec:intro}

Recently, a Lifshitz-type theory of gravity called the
Ho\v{r}ava-Lifshitz (HL) gravity~\cite{Horava:2008ih, Horava:2009uw}
has been proposed aiming at a renormalizable theory of gravity with
anisotropic scaling of space and time. The scale transformations are
defined by
\begin{equation}
  \label{z:transf:scale}
  t \to b^z t, \quad x^i \to b x^i,
\end{equation}
where $i = 1, \cdots, D$ is the spatial index, $D$ is the dimension of
space, and the Lifshitz index $z$ is the ``critical exponent'' in the
Lifshitz scalar field theory. The Lorentz invariant scale
transformation corresponds to the case of $z=1$. The ``weighted''
scaling dimension is also defined by $ [t]_w = -z$ and $[x^i]_w = -1
$. It is assumed to recover the general relativity in the IR regime whereas it
becomes a nonrelativistic gravity in the UV regime. Now, there have
been extended studies in various aspects of black
holes~\cite{Lu:2009em, Cai:2009qs, Ghodsi:2009zi, Koutsoumbas:2010pt, Koutsoumbas:2010yw, Eune:2010kx,
  Radinschi:2010iw} and cosmologies~\cite{Kiritsis:2009sh,
  Mukohyama:2009gg, Mukohyama:2009zs, Brandenberger:2009yt,
  Setare:2010wt, Jamil:2010di, Son:2010qh, Balasubramanian:2010uk,
  Saaidi:2010jq, Maeda:2010ke, Saridakis:2011pk}.  By the way, the HL gravity was
originated from a Lifshitz scalar field theory studied in the
condensed matter physics as a description of tricritical phenomena
involving spatially modulated phases~\cite{lifshitz,hls}.  Moreover,
the Lifshitz-type theory can be also studied in the framework of the
Maxwell's electromagnetic field theory~\cite{Horava:2008jf} and the
scalar field theories~\cite{Iengo:2009ix}.  Especially, in 
Ref.~\cite{Iengo:2009ix}, the one-loop renormalization and evolution
of the couplings have been studied in detail to investigate the
emergent Lorentz symmetry in various Lifshitz scalar field theories.
 
On the other hand, the effective potentials of Lorentz invariant
theories corresponding to $z=1$ have been studied through the
functional evaluation~\cite{Coleman:1973jx, Weinberg:1973ua,
  Dolan:1973qd, Dolan:1974gu, Jackiw:1974cv}. They can be widely used
in studying symmetry breaking and cosmological applications in
spite of the zero momentum limit of the effective
action~\cite{Brandenberger:1984cz}. As expected, for the case of $z=2$, the
UV divergence can be mild due to the higher-derivative Lifshitz term
which plays a role of UV-cutoff in connection with renormalization,
which eventually gives rise to the different type of effective
potential from that of the Lorentz invariant theory.  

Now, we would like to study the one-loop effective potential in the
Lifshitz scalar field theory with the anisotropic scaling of
$z=2$ in order to investigate the behaviors of the UV divergence.
The classical Lagrangian of the Lifshitz scalar takes the form of 
\begin{equation}
 \mathcal{L}_0 = \frac12 \dot\phi^2 - \frac12\alpha^2 (\partial_i^2
  \phi)^2 - \frac{m^2}{2} \phi^2 - \sum_{n=1}^{N_\lambda} \frac{\lambda_n}{(2n+2)!} 
\phi^{2n+2} - \left( \frac12 c^2 + \sum_{n=1}^{N_\eta} \frac{\eta_n}{(2n)!} \phi^{2n} \right) (\partial_i \phi)^2, \label{gen:L:0} 
\end{equation}
where it is equivalent to the action in Ref.~\cite{Iengo:2009ix} for $N_\lambda=4$ and $N_\eta=2$. 
The weighted scaling dimension of the scalar
becomes $[\phi]_w = 1/2$.  The coefficients $\alpha$, $\lambda_{N_\lambda}$ and $c$ are
assumed to be positive constants and their weighted scaling dimensions
are $[\alpha]_w = 0$, $[\lambda_n]_w = (4-n)$ and $[c]_w = 1$
so that the action is power-counting renormalizable.  In fact, the action~(\ref{gen:L:0})
becomes the well-known four-dimensional $\phi^4$-theory for $z=1$; i.e.,
$\alpha = N_\eta = 0$ and $N_\lambda = 1$ with $\lambda_1 = \lambda$.

Actually, it is not easy to get the nice closed form of the effective potential for the general action.   
So, we would like to consider the simpler case of $c=N_\eta=0$ neglecting the last term in Eq.~\eqref{gen:L:0} 
for convenience; however, the full renormalizations in the one-loop approximation will be discussed in the last section.
In section~\ref{sec:z}, the UV divergence is properly regulated in the one-loop approximation so that  the
mass and the coupling constants renormalizations are performed. 
It is interesting to note that the counterterm for the highest order of the coupling constant is \emph{finite}, 
which is in contrast with the
conventional Lorentz invariant theory.
Next, we investigate whether the critical temperature can exist or not
in~section~\ref{sec:z:T}.  Unfortunately, it turns out that the
critical temperature to recover the broken symmetry does not exist in
this approximation. Finally, conclusion will be given in section~\ref{sec:discus}. In particular,
we will discuss the counterterms for the most general action~\eqref{gen:L:0}. 

\section{Effective potential at zero temperature}
\label{sec:z}


We now study a four-dimensional Lifshitz scalar field theory of
$z=2$ which consists of the only marginally
deformed kinetic term and the full higher order of  potential terms 
instead of the most general $z=2$ theory~\eqref{gen:L:0} in order for simple
arguments. The classical Lagrangian is obtained as
\begin{equation}
  \label{z=2:L:0}
  \mathcal{L}_0^{(z=2)} = \frac12 \dot\phi^2 - \frac12\alpha^2 (\partial_i^2
  \phi)^2 - \frac{m^2}{2} \phi^2 - \sum_{n=1}^{4} \frac{\lambda_n}{(2n+2)!} \phi^{2n+2}
\end{equation}
by setting $c=N_\eta=0$ from the general action and 
the corresponding counterterms are expected as $\mathcal{L}_{ct} = - \delta m^2 \phi^2/2 - \sum_{n=1}^{N_\lambda}
 \delta \lambda_n \phi^{2n+2}/(2n+2)!$.
The $\delta m^2$ and $\delta \lambda_n$ in the counterterms
are given by power-series in $\hbar$,
\begin{align}
  \delta m^2 &= \hbar \delta m_{(1)}^2 + \hbar^2 \delta m_{(2)}^2 +
  \cdots, \label{dm} \\
  \delta \lambda_n &= \hbar \delta \lambda_n^{(1)} + \hbar^2 \delta \lambda_n^{(2)} +
  \cdots. \label{dlambda}
\end{align}
Note that we ignore the entire wave-function renormalization
counterterm in our approximation since it plays no role.

In order to calculate the effective potential on the background field
$\hat{\phi}$, we shift the field $\phi(x)$ by $\phi(x) \to \hat{\phi}
+ \varphi(x)$, and then consider the quadratic and higher orders with
respect to $\varphi(x)$. Then, the action from the
Lagrangian~(\ref{z=2:L:0}) with the counterterms becomes
\begin{equation}
  \label{z:L:trunc}
  \hat{\mathcal{L}} \{ \hat{\phi}; \varphi(x) \}=
  \hat{\mathcal{L}}_0 \{ \hat{\phi}; \varphi(x) \} +
  \hat{\mathcal{L}}_I \{ \hat{\phi}; \varphi(x) \},
\end{equation}
where 
\begin{align}
  \hat{\mathcal{L}}_0 \{ \hat{\phi}; \varphi(x) \} &= \frac12
  \varphi(x) \left[ -\partial_t^2 - \alpha^2 (\vec{\nabla}^2)^2 - \tilde{M}^2
  \right] \varphi(x), \label{z:L:1loop} \\
  \hat{\mathcal{L}}_I \{ \hat{\phi}; \varphi(x) \} &= -
  \sum_{n=1}^{4} \sum_{m=3}^{2n+2} \frac{\lambda_n \hat{\phi}^{2n+2-m}}{m!(2n+2-m)!}
  \varphi^m(x), \label{z:L:higher}
\end{align}
and $\vec{\nabla}^2$ is the spatial Laplacian and $\tilde{M}^2 \equiv
M^2 + \delta m^2 + \sum_{n=1}^4 [\delta \lambda_n/(2n)!] \hat{\phi}^{2n}$ with $M^2
\equiv m^2 + \sum_{n=1}^4 [\lambda_n/(2n)!] \hat{\phi}^{2n}$. The
$\hat{\mathcal{L}}_0$ in Eq.~(\ref{z:L:1loop}) gives one-loop
approximation and the interaction term $\hat{\mathcal{L}}_I$ in
Eq.~(\ref{z:L:higher}) contributes to higher loop calculations.  From
now on, we are going to focus on the one-loop effective potential.

The zeroth-loop effective potential is just given by the classical
form,
\begin{equation}
  \label{z:V:0loop}
  V_0(\hat{\phi}) = \frac{m^2 + \delta m^2}{2} \hat{\phi}^2 + \sum_{n=1}^4
  \frac{\lambda_n + \delta \lambda_n}{(2n+2)!} \hat{\phi}^{2n+2}.
\end{equation}
From Eq.~(\ref{z:L:1loop}), one can write down the one-loop approximation as
\begin{align}
  V_1(\hat{\phi}) &= - \frac{i\hbar}{2} \int \frac{d^4 k}{(2\pi)^4}
  \ln \left[ k_0^2 - \alpha^2 (\vec{k}^2)^2 - {M}^2
  \right]. \label{z:V1:z:4d}
\end{align}
Since the model does not have SO(4) symmetry in the Euclideanized
momentum space, which reflects the lack of the Lorentz symmetry, we
have to consider the timelike and the spacelike sectors
separately. So, the cutoff is naturally taken as a three-dimensional
momentum cutoff.  Hence, the UV cutoff, $\Lambda$, is different from the
conventional cutoff appearing in literatures.  With the help of
the following relation, apart from an infinite constant independent of
$\hat{\phi}$,
\begin{equation}
  \label{int:k0}
  -i \int_{-\infty}^{\infty} \frac{dk_0}{2\pi} \ln (k_0^2 - E^2 + i
  \epsilon) = E,
\end{equation}
when $\epsilon$ goes to zero, Eq.~(\ref{z:V1:z:4d}) becomes the
three-dimensional integral,
\begin{align}    
  V_1(\hat{\phi}) &= \frac{\hbar}{2} \int \frac{d^3 \vec{k}}{(2\pi)^3}
  {E}_M, \label{z:V1:z}
\end{align}
where the dispersion relation is nonrelativistic, ${E}_M^2 \equiv
\alpha^2 (\vec{k}^2)^2 + {M}^2$.  So, the integral~(\ref{z:V1:z})
with a UV cutoff  takes the form of
\begin{equation}
  \label{z=2:V1}
  V_1 (\hat{\phi}) = \frac{\hbar}{4\pi^2} \int_0^\Lambda dk\, k^2
  \sqrt{\alpha^2 k^4 + {M}^2},
\end{equation}
where $k$ is the magnitude of $\vec{k}$.
By integrating out the spatial momenta in Eq.~(\ref{z=2:V1}), one gets
\begin{align}
  V_1 (\hat{\phi}) &= \frac{\hbar\Lambda^3}{12\pi^2}
  (M^2)^{1/2} \, {}_2F_1 \left( -\frac12, \frac34; \frac74; - \frac{\alpha^2
    \Lambda^4}{{M}^2} \right), \label{z=2:V1:cutoff:2F1}
\end{align}
where the hypergeometric function ${}_2F_1 (a,b;c;z)$ is
defined by
\begin{equation}
  \label{def:2F1}
  {}_2F_1 (a,b;c;z) \equiv \frac{\Gamma (c)}{\Gamma (b)\Gamma (c-b)}
  \int _0^1t^{b-1}(1-t)^{c-b-1}(1-tz)^{-a} dt. 
\end{equation}
It is a solution to the differential equation, 
\begin{equation}
  \label{deom:2F1}
  z(1-z) y''+ [c-(a+b+1)z]y' - a b y = 0,
\end{equation}
so that the solution of ${}_2F_1 (a,b;c;z)$ can be expressed in terms
of series expansion of
\begin{equation}
  \label{def:2F1:series:y=0}
  {}_2F_1 (a,b;c;z) \equiv \sum_{k=0}^\infty \frac{\Gamma(a+k) \Gamma(b+k) \Gamma(c)
  }{\Gamma(a) \Gamma(b) \Gamma(c+k)} \frac{z^k}{k!}.
\end{equation}
Then, Eq.~(\ref{def:2F1}) can be also expanded as
\begin{align}
  {}_2F_1 (a,b;c;\pm z) &= (\mp z)^{-b}\left[\frac{\Gamma(a-b)
      \Gamma(c)}{\Gamma (a) \Gamma (c-b)} \pm \frac{b (1+b-c) \Gamma
      (a-b) \Gamma (c)}{(1-a+b) \Gamma (a) \Gamma (c-b)} \frac{1}{z} +
    O\left(\frac{1}{z^2}\right)\right] \notag \\
  &\quad + (\mp z)^{-a}\left[\frac{\Gamma (b-a) \Gamma (c)}{\Gamma (b)
      \Gamma (c-a)} \pm \frac{a (1+a-c) \Gamma (b-a) \Gamma
      (c)}{(1+a-b) \Gamma (b) \Gamma (c-a)} \frac{1}{z} +
    O\left(\frac{1}{z^2}\right)\right]. \label{2F1:series:z=infinity}
\end{align}
In Eq.~(\ref{2F1:series:z=infinity}), the hypergeometric function
${}_2F_1 (a,b;c;z)$ vanishes when $|z|$ goes to infinity if $a$, $b$,
$c$ are positive, $(a-b)$ is not an integer, and $c>a$ and $c>b$ are
satisfied. Using Eq.~(\ref{2F1:series:z=infinity}),
Eq.~(\ref{z=2:V1:cutoff:2F1}) can be simplified as
\begin{align}
  V_1 (\hat{\phi}) &= \frac{\hbar}{8\pi^{5/2} \alpha^{3/2}} \left[
    \sqrt{\alpha\pi} {M}^2 \Lambda - \frac45\,
    [\Gamma(3/4)]^2 ({M}^2)^{5/4} \right] + O
  \left(\frac{1}{\Lambda^2} \right).
  \label{z=2:V1:cutoff}
\end{align}
Then, for a large $\Lambda$, the effective potential
in the one-loop approximation is 
\begin{align}
  V_{\rm eff} (\hat{\phi}) &= V_0 (\hat{\phi}) + V_1
  (\hat{\phi}) \notag \\
  & \begin{aligned}[b] = \frac{m^2}{2} \hat{\phi}^2 + \sum_{n=1}^4 \frac{\lambda_n}{(2n+2)!} \hat{\phi}^{2n+2} + \frac{\hbar}{8\pi^{5/2} \alpha^{3/2}} \bigg[ & \sqrt{\alpha\pi} {M}^2 \Lambda - \frac45\, [\Gamma(3/4)]^2 ({M}^2)^{5/4} \\
    & + \frac{\delta m_{(1)}^2}{2} \hat{\phi}^2 + \sum_{n=1}^4 \frac{\delta \lambda_n^{(1)}}{(2n+2)!} \hat{\phi}^{2n+2} \bigg]. \end{aligned} \label{V:eff}
\end{align}

Now, the renormalized mass $m$ is defined by
\begin{equation}
  \label{def:mass}
  m^2 = \left. \frac{\partial^2 V_{\rm eff}}{\partial \hat{\phi}^2} \right|_{\hat{\phi}=0},
\end{equation}
from which the mass counterterm $\delta m_{(1)}^2$ can be determined in the
order of $\hbar$.
Next, to determine the counterterm $\delta \lambda_n^{(1)}$, we have to
consider the asymmetric renormalization point $M_0$ due to the IR
divergence in the massless case of $m^2=0$,
\begin{equation}
  \label{fix:dlambda1}
  \left. \frac{\partial^4 V_{\rm eff}}{\partial \hat{\phi}^4}  
  \right|_{\hat{\phi} = M_0} = \lambda,
\end{equation}
while in the massive case of $m^2\ne0$, the IR divergence does not appear so that $M_0$ can be removed.
In what follows, we will calculate the effective potential in the massless and the massive cases by using these renormalization conditions.


\subsection{$m^2=0$ case}
\label{sec:z=2:massless}
As a first application, we want to consider a massless Lifshitz scalar
field theory which is a modified Coleman-Weinberg scalar theory of
$z=1$~\cite{Coleman:1973jx}.  Before we get down to this problem, let
us consider the effective potential of $z=1$ in order to compare it
with that of $z=2$ on the same ground of the three-dimensional cutoff: the classical Lagrangian is now written as
\begin{equation}
  \label{z=1:L:0}
  \mathcal{L}_0^{(z=1)} = \frac12 \dot\phi^2 - \frac12 c^2 (\partial_i
  \phi)^2 - \frac{m^2}{2} \phi^2 - \frac{\lambda}{4!} \phi^4,
\end{equation}
and the corresponding counterterms are given by $\mathcal{L}_{ct} = - \delta m^2 \phi^2/2 - \delta \lambda \phi^{4}/4!$.
Along the usual procedure, using Eqs.~(\ref{def:mass}) and \eqref{fix:dlambda1}, the mass and the coupling constant
counterterms can be determined by
\begin{align}
  \delta m_{(1)}^2 &= - \frac{\Lambda^2 \lambda}{16 \pi^2
    c}, \label{dm1:m=0:z=1} \\
  \delta \lambda^{(1)} &= - \frac{3\lambda^2}{2^5 \pi^2 c^3} \left(
    \ln \frac{\lambda M_0^2}{8c^2 \Lambda^2} + \frac{14}{3}
      \right). \label{dlambda1:m=0:z=1}
\end{align}
If we do not consider the nontrivial renormalization point $M_0$ which
corresponds to introducing IR cutoff, we can not remove the UV
divergence.  Note that $\Lambda$ is not a cutoff defined by the
four-dimensional Euclidean length but the three-dimensional spatial
length. Actually, the counterterms are essentially the same with those
of the conventional SO(4) invariant cutoff apart from some
coefficients.  Then, the effective potential is obtained as
\begin{equation}
  \label{z=1:m=0:Veff}
     V_{\rm eff} (\hat{\phi}) = \frac{\lambda}{4!} \hat{\phi}^4 +
   \frac{\hbar \lambda^2 \hat{\phi}^4}{256\pi^2 c^3} \left[ \ln
     \frac{\hat{\phi}^2}{M_0^2} - \frac{25}{6} \right],
\end{equation}
which is exactly the same as the result for $c=1$ given in
Ref.~\cite{Coleman:1973jx, Brandenberger:1984cz}.

Now, for the case of $z=2$ with the classical action~\eqref{z=2:L:0}, one can determine the mass and the coupling constant counterterms
from the renormalization conditions~(\ref{def:mass}) and \eqref{fix:dlambda1},
\begin{align}
  \delta m_{(1)}^2 &= - \frac{\Lambda \lambda_1}{8 \pi^2 \alpha}, \label{z=2:m=0:dm1} \\
  \delta \lambda_n^{(1)} &= - \frac{\Lambda \lambda_{n+1}}{8 \pi^2 \alpha} + \tilde\lambda_n(M_0)~~(n=1,2.3,4), \label{z=2:m=0:dlambda1}
\end{align}
where the constants $\tilde\lambda_n(M_0) = [\Gamma(3/4)]^2 (10 \pi^{5/2} \alpha^{3/2})^{-1} (\partial/\partial \hat{\phi})^{2n+2} (M^2)^{5/4} |_{\hat{\phi}=M_0}$ are finite lengthy constants.
Note that we have assumed $\lambda_5 = 0$ so that $\delta \lambda_4^{(1)}$ becomes the finite counterterm.
It means that for arbitrary $N_\lambda\ge1$, the highest order counterterm $\delta \lambda_{N_\lambda}^{(1)}$ becomes independent of the UV divergence so that the improvement of the Lifshitz theory can be shown in the highest order of the coupling constant counterterm.
Then, substituting Eqs.~(\ref{z=2:m=0:dm1}) and \eqref{z=2:m=0:dlambda1} into Eq.~(\ref{V:eff}), the
renormalized effective potential can be obtained as
\begin{align}
  V_{\rm eff} (\hat{\phi}) &= \sum_{n=1}^4 \frac{(\lambda_n+\hbar\tilde\lambda_n)}{(2n+2)!} \hat{\phi}^{2n+2} -
  \frac{\hbar [\Gamma(3/4)]^2 (M^2)^{5/4}}{10 \pi^{5/2}
    \alpha^{3/2}}.
 \label{z=2:m=0:Veff:dlambda1}
\end{align}

\begin{figure}[pbt]
  \centering
  \subfigure[~The potentials at $z=1$]{
    \includegraphics[width=0.45\textwidth]{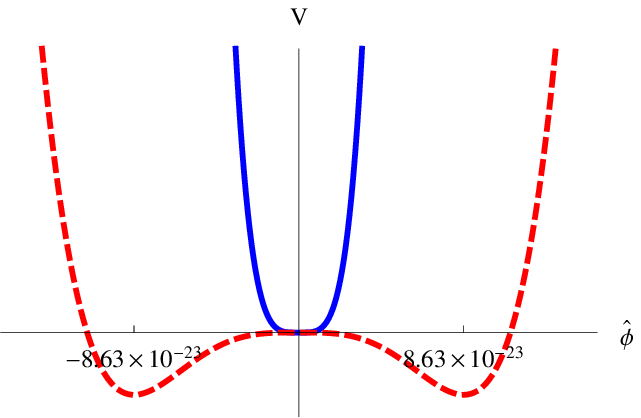} \label{fig:massless:z=1}}
  \subfigure[~The potentials at $z=2$]{
    \includegraphics[width=0.45\textwidth]{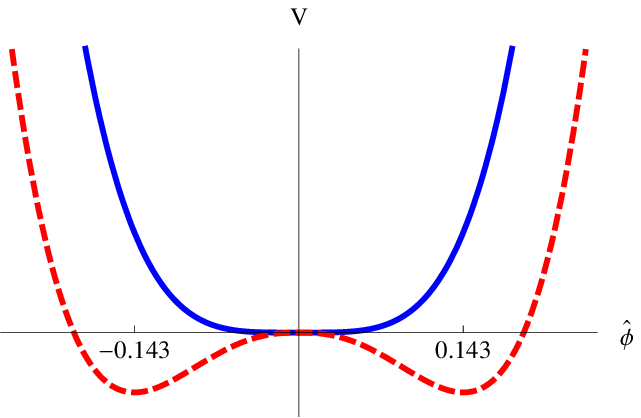} \label{fig:massless:z=2}}
  \caption{The effective potentials for $z=1$ and $z=2$ are given by
    Eq.~\eqref{z=1:m=0:Veff} and Eq.~\eqref{z=2:m=0:Veff:dlambda1},
    respectively. The solid and dashed lines represent the classical($\hbar=0$)
    and one-loop($\hbar=1$) effective potentials, respectively. The constants
    have been chosen as $N_\lambda=\lambda=\alpha=M_0 = 1$ for
    simplicity. With these constants, the effective potential has
    minimal values at $\hat{\phi} = \pm 8.62703\times10^{-23}$ and $\hat{\phi} = \pm 0.143432$ for $z=1$ and $z=2$, respectively.}
  \label{fig:massless}
\end{figure}

Basically, the effective potential is written
in terms of the (fractional) polynomials of the classical field rather
than the logarithmic type.  As for the symmetry breaking, the
effective potential~\eqref{z=2:m=0:Veff:dlambda1} shows that the
symmetry breaking still happens quantum mechanically.  Its overall
pattern is almost same with that of $z=1$ as seen from
Fig.~\ref{fig:massless}.

\subsection{$m^2  \ne 0$ case}
\label{sec:z=2:m_positive}
As was done in the previous section, we first obtain the effective potential for the case of $z=1$.
Then, the counterterms can be determined in terms of the three dimensional UV cutoff
as
\begin{align}
  \delta m_{(1)}^2 &= - \frac{\Lambda^2 \lambda}{2^4 \pi^2 c}
    - \frac{m^2\lambda}{2^5\pi^2c^3} \left( \ln \frac{m^2}{4 c^2
      \Lambda^2} + 1 \right), \label{dm1:m:z=1} \\
  \delta \lambda^{(1)} &= - \frac{3\lambda^2}{2^5 \pi^2 c^3}
    \left( \ln \frac{m^2}{4 c^2 \Lambda^2} + 1 \right), \label{dlambda1:m:z=1}
\end{align}
where we used $M_0 = 0$ in Eq.~(\ref{fix:dlambda1}) since we can avoid
the IR divergence with the help of the nonvanishing mass term.
Then, the effective potential in the Lorentz invariant scalar
field theory is
\begin{equation}
  V_{\rm eff} (\hat{\phi}) = \frac12 m^2 \hat{\phi}^2 +
  \frac{\lambda}{4!} \hat{\phi}^4 + \frac{\hbar}{64\pi^2 c^3} \left[ M^4 \ln
    \frac{M^2}{m^2} - \frac{\lambda}{2} \hat{\phi}^2 \left( m^2 + \frac34 \lambda \hat{\phi}^2 \right)
  \right], \label{V:eff:m:z=1}
\end{equation}
which agrees with the result for $c=1$ given in Refs.~\cite{Jackiw:1974cv,
  Dolan:1974gu, Weinberg:1973ua, Dolan:1973qd};
however, the potential~\eqref{V:eff:m:z=1} was shifted to satisfy $V_\text{eff}(0) = 0$.

Next, for a massive Lifshitz scalar field at a fixed point $z=2$, the
mass and the coupling constant counterterms from Eqs.~(\ref{def:mass})
and (\ref{fix:dlambda1}) can be determined as
\begin{align}
  \delta m_{(1)}^2 &= - \frac{\Lambda \lambda_1}{8\pi^2\alpha} +
    \tilde{m}^2,  \label{z=2:massive:dm1} \\
  \delta \lambda_n^{(1)} &= - \frac{\Lambda \lambda_{n+1}}{8\pi^2\alpha} + \tilde\lambda_n, \label{z=2:massive:dlambda1}
\end{align}
where $\tilde{m}^2$ and $\tilde\lambda_n$ are now defined by $\tilde{m}^2 = [\Gamma(3/4)]^2 (10 \pi^{5/2} \alpha^{3/2})^{-1} (\partial/\partial \hat{\phi})^{2} (M^2)^{5/4} |_{\hat{\phi}=0}$ and $\tilde\lambda_n = [\Gamma(3/4)]^2 (10 \pi^{5/2} \alpha^{3/2})^{-1} (\partial/\partial \hat{\phi})^{2n+2} (M^2)^{5/4} |_{\hat{\phi}=0}$.
And then, plugging Eqs.~(\ref{z=2:massive:dm1}) and \eqref{z=2:massive:dlambda1} into Eq.~(\ref{V:eff}),
the effective potential can be easily obtained as
\begin{align}
  V_{\rm eff} (\hat{\phi}) &= \frac{m^2 + \hbar \tilde{m}^2}{2} \hat{\phi}^2 +
    \sum_{n=1}^4 \frac{(\lambda_n + \hbar \tilde\lambda_n)}{(2n+2)!}  \hat{\phi}^{2n+2} -\frac{\hbar
    [\Gamma(3/4)]^2 }{10 \pi^{5/2} \alpha^{3/2}} \left[(M^2)^{5/4} - (m^2)^{5/4} \right].
  \label{z=2:massive:Veff:dlambda1}
\end{align}
Although the effective potential approximates to the classical potential with renormalized mass and coupling constants for small $\hat\phi$,  it is hard to say what happens to the effective potential  on general ground for $N_\lambda>1$.
However, simply for $m^2>0$ with $N_\lambda=1$, there is no symmetry breaking behavior.
To see this, we first calculate the
slope of Eq.~(\ref{z=2:massive:Veff:dlambda1}) written as
\begin{equation}
\partial  V_{\rm
  eff} (\hat{\phi})/\partial \hat{\phi} = \hat{\phi} f(\hat{\phi}),
\end{equation}
where $f(\hat{\phi}) 
= 2m^2/3 + 3\lambda \Delta (m^2)^{1/4}/4 + M^2/3 + \lambda^2 \Delta (m^2)^{-3/4} M^2/4 - \lambda \Delta (M^2)^{1/4}$ with a positive constant
$\Delta \equiv \hbar [\Gamma(3/4)]^2 / (2^3 \pi^{5/2} \alpha^{3/2})$.
The function $f(\hat{\phi})$ is always positive so that there is no symmetry breaking behavior.

\begin{figure}[pbt]
  \centering
  \subfigure[~The potentials at $z=1$]{
    \includegraphics[width=0.45\textwidth]{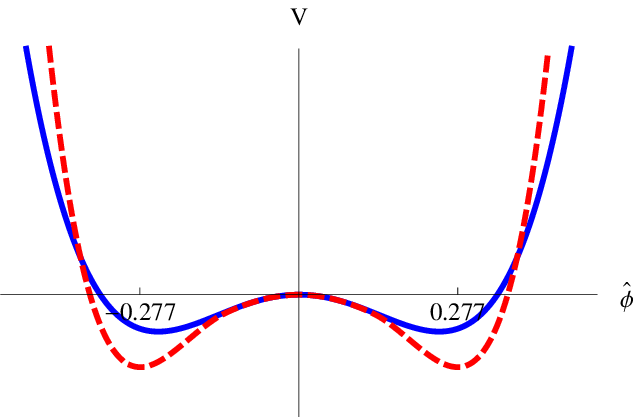} \label{fig:negative:z=1}}
  \subfigure[~The potentials at $z=2$]{
    \includegraphics[width=0.45\textwidth]{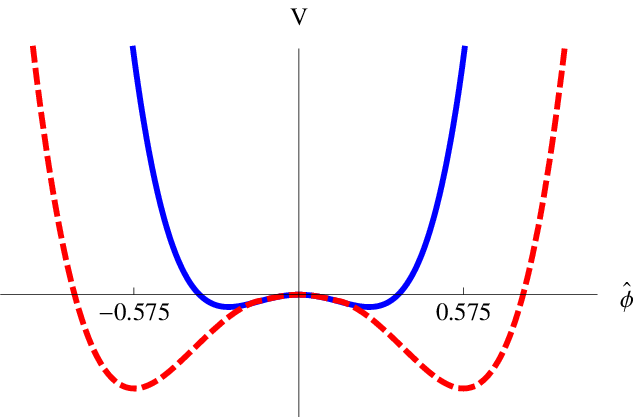} \label{fig:negative:z=2}}
  \caption{The effective potentials for $z=1$ and $z=2$ are given by
    Eq.~\eqref{V:eff:m:z=1} and Eq.~\eqref{z=2:massive:Veff:dlambda1},
    respectively. This figure shows the behavior of the effective potential
    for $m^2 < 0$. The solid and dashed lines represent the classical($\hbar=0$)
    and one-loop($\hbar=1$) effective potentials, respectively. The constants
    have been chosen as $m^2=-1$, $\lambda=100$ and $N_\lambda=\alpha= M_0 = 1$ for
    simplicity. With these constants, the effective potential has
    minimal values at $\hat{\phi} = \pm 0.276978$ and $\hat{\phi} = \pm 0.574470$ for $z=1$ and $z=2$, respectively.}
  \label{fig:negative}
\end{figure}

On the other hand, for $m^2 < 0$ corresponding to the case of the classically broken
symmetry, the effective potential may be complex depending on the
range of field for a given mass so that we have to consider the real
part of the effective potential.  Then,
Eq.~(\ref{z=2:massive:Veff:dlambda1}) can be expressed as
\begin{align}
  V_{\rm eff} (\hat{\phi}) &= \frac{m^2 + \re(\tilde{m}^2)}{2^{3/2}} \hat{\phi}^2 +
    \sum_{n=1}^4 \frac{\lambda_n + \re(\tilde\lambda_n)}{(2n+2)!}  \hat{\phi}^{2n+2} -\frac{\hbar
    [\Gamma(3/4)]^2 }{10 \pi^{5/2} \alpha^{3/2}} \left[(M^2)^{5/4} + \frac{(-m^2)^{5/4}}{\sqrt2} \right],
  \label{z=2:massive:Veff:near0}
\end{align}
for $M^2 > 0$, and
\begin{align}
  V_{\rm eff} (\hat{\phi}) &= \frac{m^2 + \re(\tilde{m}^2)}{2} \hat{\phi}^2 +
    \sum_{n=1}^4 \frac{\lambda_n + \re(\tilde\lambda_n)}{(2n+2)!}  \hat{\phi}^{2n+2} +\frac{\hbar
    [\Gamma(3/4)]^2 }{10 \sqrt2 \pi^{5/2} \alpha^{3/2}} \left[(-M^2)^{5/4} - (-m^2)^{5/4} \right],
  \label{z=2:massive:Veff:far0}
\end{align}
for $M^2 < 0$.
As seen from Fig.~\ref{fig:negative}, the
vacuum expectation values are quantum mechanically larger than the classical ones, 
in particular, remarkably for the case of $z=2$. 

\section{Effective potential at finite temperature}
\label{sec:z:T}
Now, we would like to investigate the critical temperature to give the
phase transition from the quantum-mechanically broken vacuum symmetry
to the symmetric phase.  At a finite temperature $\beta^{-1}$, the
time interval is given by $0 \le t \le - i \beta$. Then, the time
component of the four vector $k_\mu$ becomes $\omega_n = 2\pi
n/(-i\beta)$, and the effective potential~\eqref{z:V1:z:4d} can be
written as
\begin{align}
  V_1^\beta (\hat{\phi^2}) &= \frac{\hbar}{2\beta} \sum_n \int
  \frac{d^3 \vec{k}}{(2\pi)^3} \ln \left[ k_0^2 -
    \alpha^2 (\vec{k}^2)^2 - \tilde{M}^2 \right] \notag \\
  &= \frac{\hbar}{2\beta} \sum_n \int \frac{d^3 \vec{k}}{(2\pi)^3}
  \left[ - \frac{4\pi^2 n^2}{\beta^2} - \tilde{E}_M^2
  \right], \label{V1:beta:def}
\end{align}
where the summation is over $n=0, \pm 1, \pm 2, \cdots$. In order to
calculate the summation, we consider
\begin{equation}
  v(E) = \sum_n \ln \left( \frac{4\pi^2 n^2}{\beta^2} + E^2 \right). \label{v:E:def}
\end{equation}
The partial derivative of $v(E)$ with respect to $E$ is given by
\begin{align}
  \frac{\partial v(E)}{\partial E} &= \sum_n \frac{2E}{4\pi^2 n^2
    /\beta^2 + E^2} \notag \\
  &= 2\beta \left( \frac12 + \frac{1}{e^{\beta E} - 1}
  \right), \label{dv:dE}
\end{align}
where the second line has been obtained using the equality $
\sum_{n=1}^\infty \frac{y}{y^2 + n^2} = - \frac{1}{2y} + \frac{\pi}{2}
\coth \pi y$.  Integrating out Eq.~(\ref{dv:dE}) with respect to $E$,
we obtain
\begin{equation}
  v(E) = 2\beta \left[ \frac{E}{2} + \frac{1}{\beta} \ln \left( 1 -
      e^{-\beta E} \right) \right] + \mathrm{const.} \label{v:E:result}
\end{equation}
As a result, the effective potential at the finite temperature in
order of $\hbar$,
\begin{align}
  V_1^\beta (\hat{\phi^2}) &= \hbar \int \frac{d^3 \vec{k}}{(2\pi)^3}
  \left[ \frac{E_M}{2} + \frac{1}{\beta} \ln \left( 1 -
      e^{-\beta E_M} \right) \right] \notag \\
  &= V_1^0 (\hat{\phi^2}) + \bar{V}_1^\beta
  (\hat{\phi^2}), \label{V1:beta:result}
\end{align}
where the zero-temperature one-loop term $V_1^0 (\hat{\phi^2})$ and
the temperature-dependent one-loop term $\bar{V}_1^\beta
(\hat{\phi^2})$ are
\begin{align}
  V_1^0 (\hat{\phi^2}) &= \hbar \int \frac{d^3 \vec{k}}{(2\pi)^3}
  \frac{E_M}{2} = \frac{\hbar}{2} \int \frac{d^3 \vec{k}}{(2\pi)^3}
  \sqrt{\alpha^2 (\vec{k}^2)^2 +
    M^2}, \label{V1:1loop:0T} \\
  \bar{V}_1^\beta (\hat{\phi^2}) &= \frac{\hbar}{\beta} \int \frac{d^3
    \vec{k}}{(2\pi)^3} \ln \left( 1 - e^{-\beta E_M} \right)
  \notag \\
  &= \frac{\hbar}{2\pi^2 \beta} \int_0^\infty dk \, k^2 \ln \left( 1 -
    e^{-\alpha\beta\sqrt{k^4 + a^2}} \right), \label{V1:1loop:T}
\end{align}
with $a^2 \equiv M^2/\alpha^2$.  Then, the critical temperature
$\beta_c$ which recovers the symmetry can be determined by~\cite{Dolan:1974gu}
\begin{align}
  - \frac{m^2}{2} &= \left.\frac{\partial
      \bar{V}_1^{\beta_c}}{\partial \hat{\phi}^2}
  \right|_{\hat{\phi}=0} \notag \\
  &= \frac{\hbar \lambda}{8\pi^2 \alpha} \int_0^\infty dk
  \frac{k^2}{\sqrt{k^4 + m^2/\alpha^2} \left(
      e^{\alpha\beta_c\sqrt{k^4 + m^2/\alpha^2}} - 1
    \right)}. \label{critical:condition}
\end{align}
Note that there is no critical temperature for $m^2>0$ since the right
hand side in Eq.~(\ref{critical:condition}) is always positive due to
the positivity of the integrand for the whole range.  For $m^2 =0$,
the integrand is divergent. In particular, for $m^2<0$, the integral
in Eq.~(\ref{critical:condition}) would lead to complex values which
are unphysical. However, we can avoid the complex values by
restricting the range of momentum in Eq.~(\ref{critical:condition}) by
setting the lower bound as $\epsilon \equiv
(-m^2/\alpha^2)^{1/4}$. Unfortunately, the integral is divergent, which means
that there is no critical temperature in the one-loop approximation.

\section{Discussion}
\label{sec:discus}
We have studied the four dimensional Lifshitz scalar field theory of
the anisotropic scaling of $z=2$, and obtained the renormalized
one-loop effective potential. The UV divergence is slightly
improved in that the finite counterterm is needed only
for the highest order of the coupling constant.
For $m^2>0$ with $N_\lambda=1$, there is no symmetry breaking behavior.
For $m^2 <0$ , the overall behavior of the
effective potential of $z=2$ is analogous to that of $z=1$; however, the vacuum expectation value is significantly larger than the classical vacuum expectation value.
Unfortunately, the critical temperatures can not be obtained in this
one-loop approximation.

On the other hand, if one considers the most general action of Eq.~\eqref{gen:L:0} 
with the counterterms given by $\mathcal{L}_{ct} = - \delta m^2 \phi^2/2 - \sum_{n=1}^{N_\lambda}
 \delta \lambda_n \phi^{2n+2}/(2n+2)!$,
the mass and the coupling constant counterterms are calculated as
$\delta m_{(1)}^2 = - \Lambda^3 \eta_1/(24\pi^2\alpha) - \Lambda (4\alpha^2\lambda_1-c^2\eta_1)/(32\pi^2\alpha^3) + \tilde{m}^2$, $\delta \lambda_1^{(1)} = - \Lambda^3 \eta_2/(24\pi^2\alpha) - \Lambda (48\alpha^2\lambda_2-3\eta_1^2-c^2\eta_2)/(3\cdot2^7\pi^2\alpha^3) + \tilde\lambda_1$, $\delta \lambda_2^{(1)} = - \Lambda (2\alpha^2\lambda_2-15\eta_1\eta_2)/(16\pi^2\alpha^3) + \tilde\lambda_2$, $\delta \lambda_3^{(1)} = - \Lambda (2\alpha^2\lambda_2-35\eta_2^2)/(16\pi^2\alpha^3) + \tilde\lambda_3$, and $\delta \lambda_4^{(1)} = \tilde\lambda_4$,
where $\tilde{m}^2$ and $\tilde{\lambda}_n$'s are just finite constants.
Unfortunately, we can not exhibit the effective potential, $\tilde{m}^2$ and $\tilde{\lambda}_n$ explicitly 
because they are so lengthy. However, it is interesting to note that the highest order of coupling constant counterterm $\delta\lambda_4^{(1)}$ is still independent of the UV cutoff.
Furthermore, taking the limit of $c\to0$ and $\eta_n\to0$, the above counterterms become exactly the same as Eqs.~\eqref{z=2:m=0:dm1} and \eqref{z=2:m=0:dlambda1}, while they are not compatible with those of the $z=1$ case because the limit of $\alpha\to0$ is ill-defined.

\begin{acknowledgments}
  M.~Eune was
  supported by National Research Foundation of Korea Grant funded by
  the Korean Government (Ministry of Education, Science and
  Technology) (NRF-2010-359-C00007).  W.~Kim and E.~J.~Son were supported by the National
  Research Foundation of Korea(NRF) grant funded by the Korea
  government(MEST) through the Center for Quantum Spacetime(CQUeST) of
  Sogang University with grant number 2005-0049409, and W.~Kim was also supported by the
  Basic Science Research Program through the National Research
  Foundation of Korea(NRF) funded by the Ministry of Education,
  Science and Technology(2010-0008359).
\end{acknowledgments}



\end{document}